\documentclass[aps,prd,showpacs,amssymb,superscriptaddress, notitlepage,floats,floatfix,nofootinbib, twocolumn,groupedaddress, longbibliography]{revtex4-2}

\usepackage{dcolumn}
\usepackage{bm}

\usepackage{amsmath}	
\usepackage{amssymb}
\usepackage[utf8]{inputenc}
\usepackage{graphicx}
\usepackage{verbatim}
\usepackage{spverbatim}
\usepackage{float}
\usepackage[section]{placeins}
\usepackage[dvipsnames]{xcolor}
\usepackage{soul}
\usepackage{xfrac}
\usepackage{changepage}
\usepackage{mathtools, wasysym}
\usepackage{relsize}
\usepackage{amsmath,tabularx,booktabs}
\usepackage{ragged2e}
\usepackage[font=small,justification=raggedright]{caption}
\usepackage[font=small,justification=justified]{subcaption}
\newcolumntype{C}{>{\Centering\hspace{0pt}}X}

\newcommand\mytab[1]{
    \smash{\begin{tabular}[t]{@{}>{\Centering}p{\hsize}@{}}
    \mdseries #1 \end{tabular}}}
    
    \newcommand{\gtsima}{$\; \buildrel > \over \sim \;$}
\newcommand{\ltsima}{$\; \buildrel < \over \sim \;$}
\newcommand{\prosima}{$\; \buildrel \propto \over \sim \;$}
\newcommand{\gsim}{\lower.5ex\hbox{\consistegtsima}}
\newcommand{\lsim}{\lower.5ex\hbox{\ltsima}}
\newcommand{\simgt}{\lower.5ex\hbox{\gtsima}}
\newcommand{\simlt}{\lower.5ex\hbox{\ltsima}}
\newcommand{\simpr}{\lower.5ex\hbox{\prosima}}
\newcommand{\sgrA}{Sgr\,A$^\star$}
\newcommand{\mpbh}{m_{\textrm{\scalebox{.75}{PBH}}}}
\newcommand{\lw}{\scalebox{.55}}
\newcommand{\comm}[1]{}

\usepackage{array}
\newcolumntype{x}[1]{>{\centering\let\newline\\\arraybackslash\hspace{0pt}}p{#1}}
\newcolumntype{y}[1]{>{\raggedright\let\newline\\\arraybackslash\hspace{0pt}}p{#1}}

\begin{document}
\title{On the detectability of gravitational waves from primordial black holes orbiting \sgrA}
\author{Stefano Bondani$^1$}
\author{Francesco Haardt$^{1,2}$}
\author{Alberto Sesana$^{3,2}$}
\author{Enrico Barausse$^{4}$}
\author{Massimo Dotti$^{3,2}$}
\affiliation{$^1$DiSAT, Universit\`a degli Studi dell'Insubria, Como, Italy}
\affiliation{$^2$INFN, Sezione Milano-Bicocca, Milano, Italy}
\affiliation{$^3$Dipartimento di Fisica, Universit\`a degli Studi di Milano-Bicocca, Milano, Italy}
\affiliation{$^4$SISSA, Trieste, Italy}
\date{\today}

\begin{abstract}
\noindent Primordial black holes, allegedly formed in the very early Universe, have been proposed as a possible viable dark matter candidate.
In this work we characterize the expected gravitational wave signal detectable by the planned space-borne interferometer LISA and the proposed next generation space-borne interferometer $\mu$Ares arising from a population of primordial black holes orbiting \sgrA, the super-massive black hole at the Galactic center. Assuming that such objects indeed form the entire diffuse mass allowed by the observed orbits of stars in the Galactic center ($\lsim 4 \times 10^3$\,M$_{\odot}$ within a radius of $\simeq 10^{-3}$\,pc from \sgrA), under the simplified assumption of circular orbits and monochromatic mass function, we assess the expected signal in gravitational waves, either from resolved and non-resolved sources. We estimate a small but non negligible chance of $\simeq 10$\% of detecting one single 1\,M$_{\odot}$ primordial black hole with LISA in a 10-year-long data stream, while the background signal due to unresolved sources would essentially elude any reasonable chance of detection. On the contrary, $\mu$Ares, with a $\simeq 3$ orders-of-magnitude better sensitivity at $\simeq 10^{-5}$\,Hz, would be able to resolve $\simeq\,140$ solar mass primordial black holes in the same amount of time, while the unresolved background should be  observable with an integrated signal-to-noise ratio $\gtrsim 100$. Allowing the typical PBH mass to be in the range 0.01-10\,M$_{\odot}$ would increase LISA chance of detection to $\simeq 40$\% towards the lower limit of the mass spectrum. In the case of $\mu$Ares, instead, we find a ``sweet spot" just about 1\,M$_{\odot}$, a mass for which the number of resolvable events is indeed maximized.
\end{abstract}
\maketitle
\section{\label{sec_intro}introduction}
In recent years, partly motivated by the inconclusive results of many enterprises aimed at the detection of dark-matter particles (for a review see, e.g., \cite{reviewdm, directdm1, directdm2}), primordial black holes (PBHs) gained increasing attention as a possible candidate \cite{carr1, PBHasDM, Green_2021, hints} for such an elusive component, which accounts to 25\% of the energy density of today's Universe. 

The existence of PBHs as physical objects was first proposed in 1966 by Zeldovich \cite{zeldovic66}, and in 1971 Hawking similarly postulated how such objects could originate \cite{hawking}. Current models trace the origin of PBHs to the collapse of large density perturbations in the early Universe, usually in the post-inflation era $t\gtrsim 10^{-38}$\,s \cite{sasaki, sasaki99, hawke, Musco2005, Musco2009}. More recent work on possible formation channels include: formation by isocurvature perturbations, such as fragmentation of a real scalar field \cite{Cotner19}, resonant amplification of the curvature perturbations \cite{Zhou20}, vacuum tunneling during inflation \cite{Deng17}, and scalaron+$\chi$ models \cite{Pi18}, among others.

Ref. \cite{carrhawking}  first proposed a simple relation between the typical mass of a newly formed PBH $m_{\scalebox{.55}{PBH}}$ and its formation cosmic time $t$, i.e.,
\begin{equation}
    \mpbh\simeq\frac{c^3t}{G}\simeq10^{15}\Bigl(\frac{t}{10^{-23}\,\textrm{s}}\Bigr)\textrm{g}\simeq10^5\Bigl(\frac{t}{1\,\textrm{s}}\Bigr)M_{\odot}.
\end{equation}
Accordingly, since black holes lose mass because of Hawking radiation \cite{hawking}, a PBH is expected to evaporate completely in a timescale given by \cite{carr2021}:
\begin{equation}
         \tau\simeq\frac{G^2\mpbh^3}{\hbar c^4}\simeq10^{64}\Bigl(\frac{\mpbh}{M_{\odot}}\Bigr)^3\,\textrm{yrs}.
\end{equation}
 While a lower limit on the current mass of a PBH is obtained by setting the evaporation time equal to the Hubble time, i.e., $\mpbh \simgt 10^{15}$\,g, no proper upper limits exist, at least on a theoretical ground. It is worth mentioning that recent theoretical arguments by \cite{prefmass} suggest that, under the assumption of a scale-invariant amplitude of primordial curvature fluctuations, the resulting PBH mass spectrum should show a clear peak at $\simeq 1\,M_{\odot}$. 

In terms of energy density, a first order estimate of the current contribution of PBHs to the dark matter component of the Universe is given in \cite{carr2021} as
\begin{equation}
    f_{\lw{PBH}}\equiv \frac{\Omega_{\lw{PBH}}}{\Omega_{\lw{CDM}}}\simeq\Bigl(\frac{\beta}{10^{-18}}\Bigr)\Bigl(\frac{\mpbh}{10^{15}\,\textrm{g}}\Bigr)^{-1/2},
\end{equation}
where $\Omega_{\lw{PBH}}$ and $\Omega_{\lw{CDM}}$ are the current density parameters in PBHs and in cold dark-matter, respectively, while $\beta$ is the fraction of the Universe mass in PBHs at their formation time. 
A number of different techniques aimed at determining upper limits on the PBH fraction as a dark matter component have been proposed, namely gravitational lensing, dynamical effects, influence on large-scale structure, accretion and gravitational waves (see \cite{carr2021} for a recent review). In particular, \cite{wang_2020} estimated the properties of a population of PBHs orbiting the supermassive variety of BHs at the center of galaxies, deriving the expected signal in gravitational waves (GWs). After limiting the analysis to the case of \sgrA, thanks to existing scaling relations \cite{Ferrarese2000,Tremaine2002} the authors of \cite{wang_2020} assessed the GW signal arising from the entire cosmic population of galaxies. Further work on the detectability of a GW stochastic background from PBHs can be found in \cite{GW_BGfromPBH, GW_BGfromPBH2}.

In the present paper we extend and refine the analysis of \cite{wang_2020}, by improving the physical soundness of the model in many aspects. In Section II we compute the expected density profile for PBHs orbiting \sgrA ~near the innermost stable circular orbit (ISCO), by considering the combined effects of two-body relaxation {\it and} GW losses, anchoring the PBH population to the one key observational constraint given by GRAVITY \cite{gravity21}. In particular, we adopt the upper limit of the diffuse mass allowed within the pericenter of the S2 star around \sgrA, i.e., $\lesssim 4\times10^3$\,M$_{\odot}$ within $r_{\scalebox{.55}{S2}}\equiv 6\times 10^{-4}$\,{pc} from \sgrA ~(i.e., about 1400 Schwarzschild Radii). In Section III we present a brief outline of the basic theoretical background of GW detection. In Section IV, through dedicated, extensive Monte-Carlo simulations, we estimate the GW characteristic strain from such constrained population of PBHs, considering observations performed by the planned space-borne interferometer LISA \cite{2017arXiv170200786A} and by the proposed next generation space-borne interferometer $\mu$Ares \cite{muares}. A distinction between resolved events and stochastic background is also then carried out before calculating the corresponding signal to noise ratios. Finally, Section V is dedicated to concluding remarks. 

\section{\label{sec_density}density distribution of PBHs around \sgrA}
\begin{figure}[t]
\vspace{-0.4cm}
\includegraphics[scale=0.55]{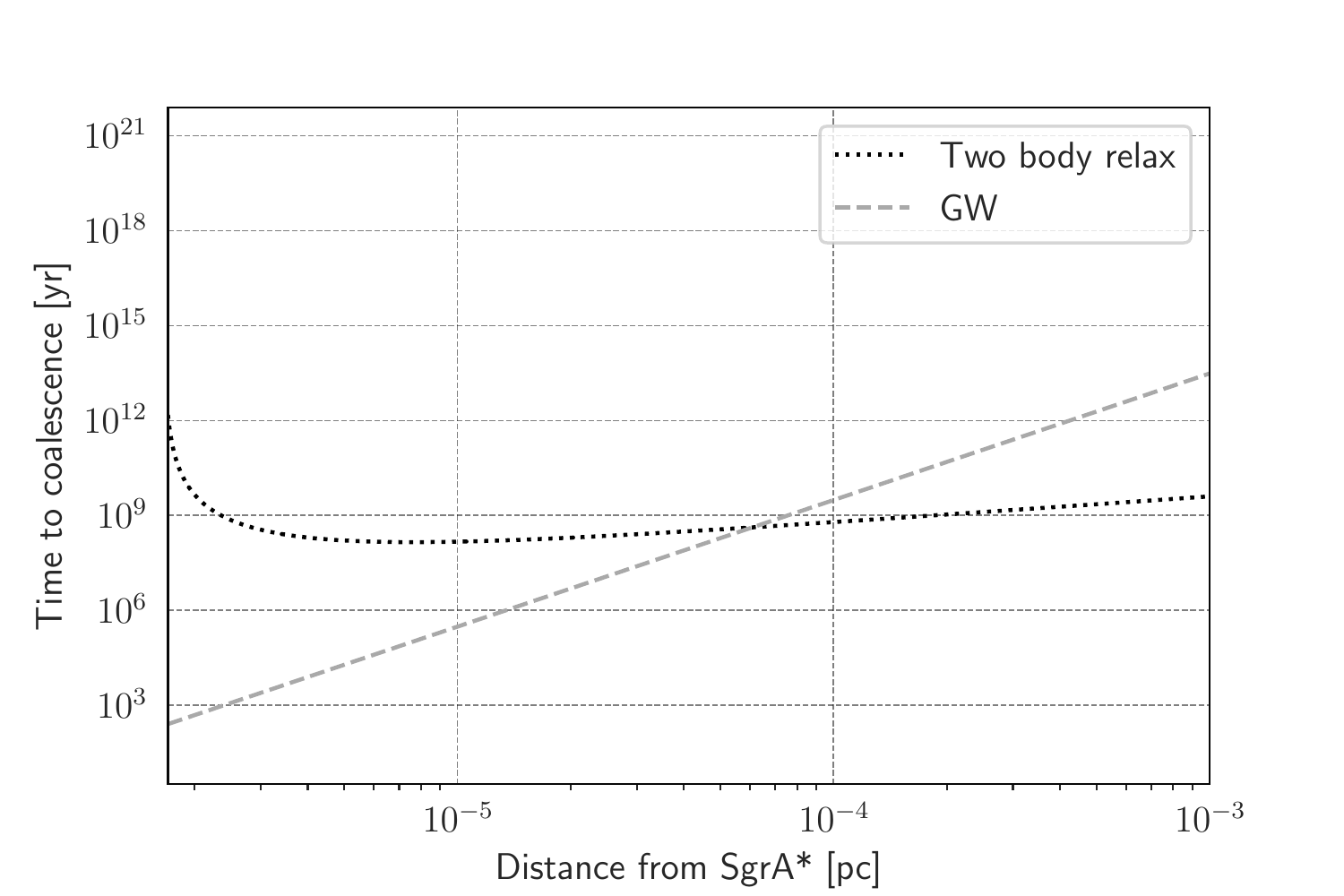}
\caption{\raggedright Characteristic timescales of 2-body relaxation (dotted line)  and GW orbital decay (dashed line) for $\mpbh=1$\,M$_\odot$, from the ISCO of \sgrA \,out to 10$^{-3}$\,pc. The small kink in the first of the two is due to the peculiar shape of the density profile from \cite{wang_2020}.}
\label{timescales}
\end{figure}

\begin{figure}
\vspace{-0.515cm}
{\includegraphics[scale=0.5]{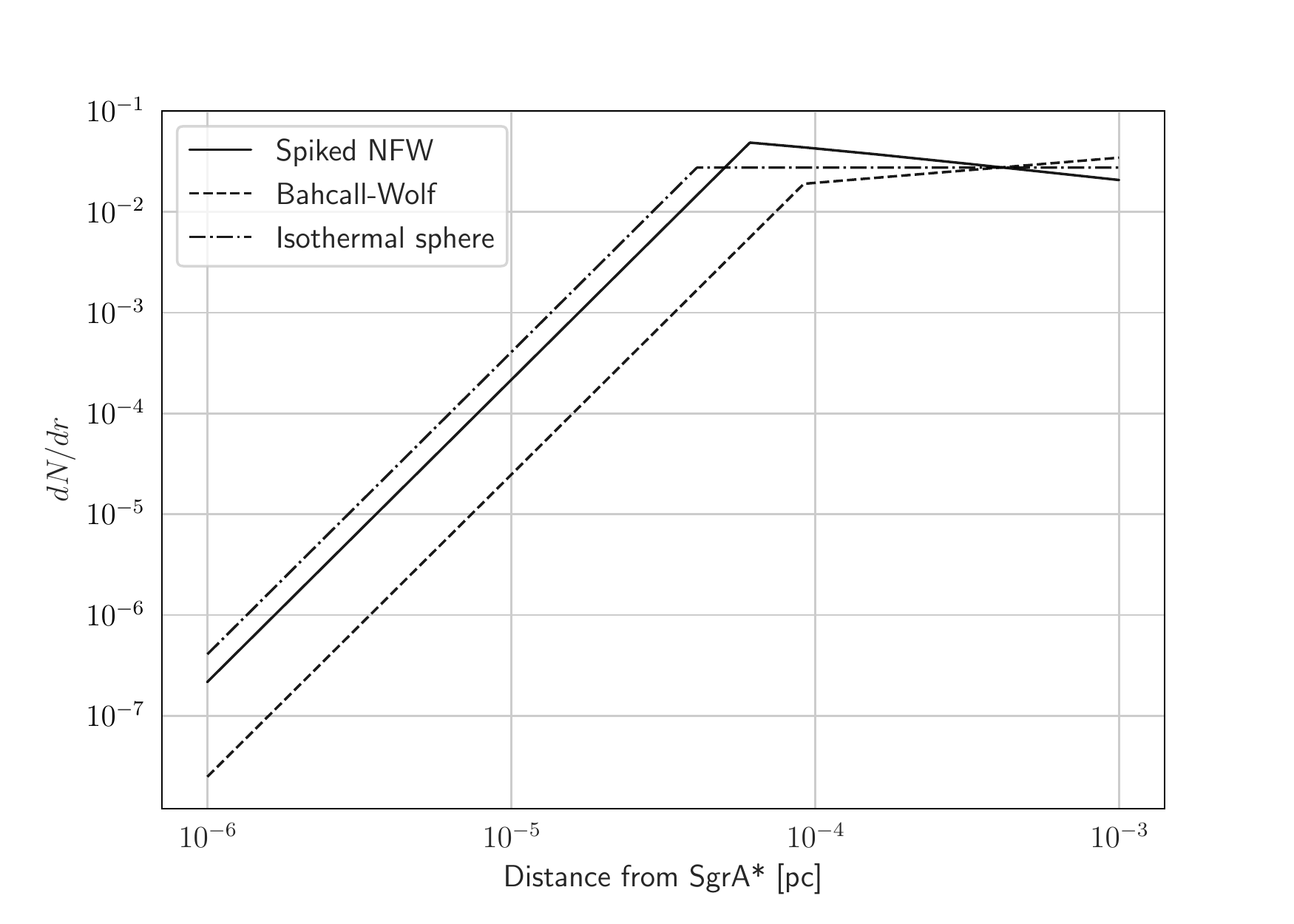}} 
\caption{\raggedright Number density distribution $dN/dr$ of 1\,M$_{\odot}$ PBHs around \sgrA, with GW sink, showing the number of PBHs in a spherical shell between $r$ and $r+dr$. The normalization is for a total mass of $4\times10^{3}$\,M$_{\odot}$. Curves refer to spiked NFW profile (solid line), Bahcall-Wolf profile (dashed line), and isothermal profile (dot-dashed line).}
\label{rhoGW2br}
\end{figure}

As in \cite{wang_2020}, we assume that a population of PBHs with typical mass $m_{\scalebox{.55}{PBH}}= 1\, M_{\odot}$ constitutes a fraction of the dark matter in the galactic center. As already pointed out, recent theoretical developments by \cite{prefmass} indicate indeed $1\,M_{\odot}$ as the preferred mass of PBHs (still, we will relax the assumption on the mass later on). PBHs are assumed to orbit \sgrA ~on purely circular orbits, and are supposed to be initially distributed according to a Navarro-Frenk-White (NFW) profile \cite{nfw}. Two-body relaxation then shapes the density profile $\rho(r)$ on a characteristic timescale given by \cite{binney}
\begin{equation}
\tau_{\scalebox{.55}{2BR}} = \frac{1.8\times10^{10}\,\textrm{y}}{\log(M_{\scalebox{.55}{MBH}}/m_{\textrm{\scalebox{.75}{PBH}}})}
\frac{1M_{\odot}}{m_{\textrm{\scalebox{.75}{PBH}}}}\frac{10^{3} M_{\odot}\textrm{pc}^{-3}}{\rho(r)}\left(\frac{v(r)}{10 {\rm km\,s^{-1}}}\right)^{3},
\label{eq:t2br}
\end{equation}
where $M_{\scalebox{.55}{MBH}}$ is the mass of the central black hole, $m_{\textrm{\scalebox{.75}{PBH}}}$ is the typical mass of PBHs, and $v(r)$ is the Keplerian mean velocity $\sqrt{(G M_{\scalebox{.55}{MBH}}/r)}$. Adopting $M_{\scalebox{.55}{MBH}}=4.3\times10^6\,M_{\odot}$ \cite{gravity21}, and by assuming the PBHs to be distributed within a spherical shell comprised between $10^{-6}$\,pc and $10^{-3}$\,pc from \sgrA, we found a maximum mass density of PBHs of few $10^{15} M_{\odot}$\,pc$^{-3}$ at $\simeq3\times10^{-6}$\,pc distance from \sgrA, in agreement with \cite{wang_2020}.
To be more precise, over a relaxation time, a spike with $\rho\propto r^{-7/3}$ is expected to form, and this is the default model that we consider here (with a cautionary $\gamma=1$ power index in Eq.\,1 of \cite{wang_2020}). In the remainder of this work we will refer to this as a Spiked NFW profile.
However, though neglected by \cite{wang_2020}, GW-driven inspirals and plunges of PBHs onto \sgrA ~largely deplete the PBH population in the very center on a characteristic timescale given by \cite{Maggiore2007}
\begin{equation}
\tau_{\scalebox{.55}{GW}}=\frac{5}{256}\frac{c^{5}r^{4}}{G^{3}m_{\scalebox{.55}{12}}^{2}\mu}\simeq 
4\times 10^{12}\,\textrm{y}\, \left(\frac{r}{r_{\scalebox{.55}{S2}}}\right)^4,
\label{eq:tGW}
\end{equation}
where $m_{\scalebox{.55}{12}}$ is the total mass of the binary and $\mu$ is the reduced mass. The numerical value is appropriate for a $1\,M_{\odot}$ PBH orbiting \sgrA ~at the S2 pericenter.
The above equation assumes circular orbits and quadrupole approximation. The two timescales $\tau_{\scalebox{.55}{2BR}}$ and $\tau_{\scalebox{.55}{GW}}$ are shown as functions of the distance from \sgrA \,in Fig.\,\ref{timescales} as dotted and dashed lines, respectively, for the aforementioned selection of the input parameters. It is apparent how closer to \sgrA \,than $\simeq\,6\times10^{-5}$\,pc the dynamical evolution of the PBHs population is no longer relaxation-dominated, but is  driven by GW radiation losses. At such characteristic distance the time to coalescence because of GWs is $\simeq 4.2\times10^{8}$\,y. Finally, the PBH number density distribution $dN/dr$ (i.e., the number of objects within $r$ and $r+dr$ distance from \sgrA) can  be found by combining these two processes, and the resulting steady-state PBH number density profile, normalized so that the diffused mass within $\simeq 6\times 10^{-4}$\,pc from \sgrA ~is $\lsim 4\times10^{3} M_{\odot}$, is displayed in Fig.~\ref{rhoGW2br} for a number of plausible relaxation-driven density profiles. In the GW-domain region the PBH number density is $\propto r^4$, while in the outer relaxation-domain it will follow the assumed density profile. In this work we first consider the Spiked NFW density profile (solid line in Fig.~\ref{rhoGW2br}) from \cite{gondolo, wang_2020}, while in Section\,\ref{sec_results} we further study the cases for a Bahcall-Wolf (dashed line) and isothermal sphere (dot-dashed line) density profiles.
It is interesting to note how, given the similar scaling with $m_{\textrm{\scalebox{.75}{PBH}}}$ of $\tau_{\scalebox{.55}{2BR}}$ and $\tau_{\scalebox{.55}{GW}}$ (Eq.~\ref{eq:t2br} and Eq.~\ref{eq:tGW}, respectively), the location of the turning point in the distribution shown in Fig.~\ref{rhoGW2br} is almost independent of the actual value of $\mpbh$. 
\section{\label{sec_results}gravitational wave signals}
In this section we introduce the observables we consider in order to characterize the GW signal arising from the population of PBHs described in the previous sections.

Both the frequency dependent strain amplitude $h(f)$ and the interferometer sensitivity $S_n(f)$ in general depend upon the position of the GW source in the sky relative to the detector. It is then useful to make the distinction between an “optimal case" and an “average case", in terms of the orientation of the source-detector system. Following, e.g., \cite{maggiore2000}, for an interferometer with non-perpendicular arms such as LISA, letting $\alpha$ be the angle between two arms, the detector beam pattern functions $F_+$ and $F_{\times}$ are defined as
\begin{align}
    F_+(\theta,\phi,\psi)=\sin\alpha \Big[\frac{1}{2}(1+\cos^2\theta)\sin(\alpha+2\phi)\cos2\psi\\
    +\cos\theta\cos(\alpha+2\phi)\sin2\psi\Big]\notag,
\end{align}
\begin{align}
    F_{\times}(\theta,\phi,\psi)=\sin\alpha \Big[\frac{1}{2}(1+\cos^2\theta)\sin(\alpha+2\phi)\sin2\psi\\
    -\cos\theta\cos(\alpha+2\phi)\cos2\psi\Big]\notag,
\end{align}
where the angles $\theta$ and $\phi$ define the source sky position, and $\psi$ is the angle of the polarization plane with respect to the line of sight. The strain amplitude in the time domain $h(t)$ is then
\begin{equation}
h(t)=F_+(\theta,\phi,\psi)h_+(t)+F_{\times}(\theta,\phi,\psi)h_{\times}(t),
\end{equation} 
where, in the case of circular orbits, the amplitudes for the two polarization modes are \cite{Maggiore2007}
\begin{align}
    h_+(t)=\frac{4}{r}\Bigl(\frac{G\mathcal{M}_{c}}{c^2}\Bigr)^{5/3}\Bigl(\frac{\pi f}{c}\Bigr)^{2/3}\Bigl(\frac{1+\cos^2\iota}{2}\Bigr)\\ \times\cos(2\pi f\,t_{\textrm{ret}}+2\phi_0)\notag,
\end{align}
\begin{align}h_{\times}(t)=\frac{4}{r}\Bigl(\frac{G\mathcal{M}_{c}}{c^2}\Bigr)^{5/3}\Bigl(\frac{\pi f}{c}\Bigr)^{2/3}\cos\iota \\ \times\sin(2\pi f\,t_{\textrm{ret}}+2\phi_0)\notag.
\end{align}
Here $t_{\textrm{ret}} = t-\frac{\lvert \textbf{x}-\textbf{x$^{\prime}$}\lvert}{c}$ indicates retarded time, where $\textbf{x}$ and $\textbf{x$^{\prime}$}$, as per the definition in Chapter 3 of \cite{Maggiore2007} are, respectively, the distances at any time $t$ from the source's center to the detector and to a point inside the source.

The “optimal orientation" case occurs when the source has angular momentum directed towards the detector, so it is face on, e.g., when $\theta,\phi$ and $\psi$ combine to give  
\begin{eqnarray}
F_+&=1,\\ 
F_{\times}&=0,
\end{eqnarray}
while the “average case" is defined whenever
\begin{equation}
    \langle F_+^2\rangle=\langle F_{\times}^2\rangle.
\end{equation}
For further reference, see \cite{maggiore2000} and \cite{hawking300}, or also \cite{2021arXiv210801167B}.
In our work we will make use of the inclination-and-polarization averaged strain, i.e., 
\begin{equation}
h=\sqrt{\langle h_+^2+h_{\times}^2\rangle}.
\end{equation}
{Since we are considering sources in the Galactic center, we can ignore factors $(1+z)$ stemming from the Universe expansion.} The inclination-and-polarization averaged strain amplitude $h$ as a function of frequency is {therefore simply} given by \cite{sesa08}
\begin{equation}
h(f,d) = \frac{8\pi^{2/3}G^{5/3}f^{2/3}\mathcal{M}_{c}^{5/3}}{c^{4}\sqrt{10}\,d},
\end{equation}
where $\mathcal{M}_{c}=\mu^{3/5}m_{\scalebox{.55}{12}}^{2/5}$ is the chirp mass and $d$ is the distance to the source. From the strain, the signal to noise ratio of the event can be computed as 
\begin{equation}
    {\rm SNR}(f) = \frac{h(f)\sqrt{\mathcal{N_{\textrm{\scalebox{.7}{cyc}}}}(f)}}{\sqrt{f S_n(f)}},
    \label{eq:SNR}
\end{equation}
where, assuming that binaries are quasi-stationary during the observation time, the number of cycles as a function of observation time $t_{\scalebox{.7}{obs}}$ and frequency $f$, is given by $\mathcal{N}_{\scalebox{.7}{cyc}}=t_{\scalebox{.7}{obs}}\times f$.
In Eq.\,\ref{eq:SNR}, the so-called noise spectral density $S_{n}(f)$ (sometimes referred to as the noise spectral sensitivity or spectral amplitude) has units [\small{Hz$^{-1}$}\normalsize], and quantifies the sensitivity of the GW detector.
\begin{figure}[ht]
\vspace{-0.35cm}
\centering

{\includegraphics[scale=0.52]{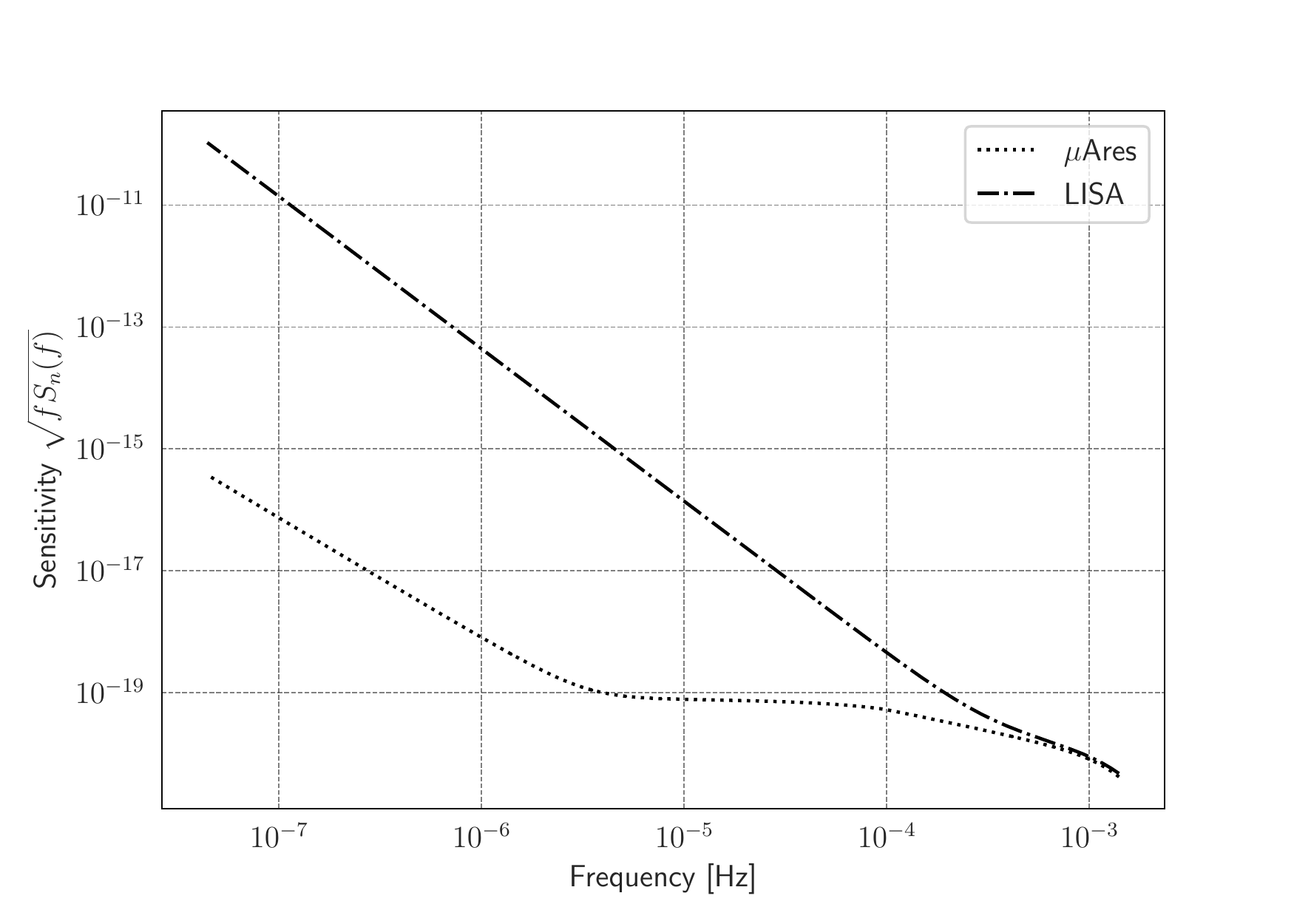}}
\vspace{-0.5cm}\caption{Sensitivities for LISA (dot-dashed line) and $\mu$Ares (dotted line) in the relevant frequency range.}
\label{cfr_sens}
\end{figure}
When computing the expected GW signal detectable by LISA we will adopt the noise spectral density reported in \cite{lisa}. Although the sources that we consider are originated at the Galactic center, we will use the sky averaged $S_n(f)$. While this choice is primarily driven by simplicity, it should be noted that PBHs are persistent sources and their signal will build-up in the data stream for the whole duration of the mission. Being the north ecliptic pole tilted by approximately 60$^{\rm o}$ with respect to the galactic north pole and being the LISA constellation tilted by 60$^{\rm o}$ with respect to the ecliptic, along the LISA orbit, the galactic center will be seen at a variable inclination, spanning a wide range essentially from being face-on to being edge-on. As for $\mu$Ares, the proposed design features two constellations in perpendicular planes, making the use of sky-averaged sensitivity a reasonable compromise in both cases.

Finally, we will add to the instrumental sensitivity curve the background noise arising from the cosmic population of white dwarf (WD) binaries \cite{wd,wdcornish}. When forecasting the GW signal in $\mu$Ares data stream, we will take $S_{n}(f)$ from \cite{muares}; note that in this case the instrumental noise already takes into account the WD background (see also \cite{lunarlaser} for a recent take on an alternative technique to detect stochastic backgrounds in the $\mu$Hz band relying on binary resonance probes).

The observation time $t_{\scalebox{.7}{obs}}$ (i.e., the proper ``data-taking" time) for the LISA interferometer is currently set to be $\gtrsim 4.5\,$yrs long, though potentially 10 years of data could be collected, if mission operations are extended and there is no failure in the hardware. Similarly, for the proposed $\mu$Ares mission, at the time of writing a $\simgt$10-year long mission is foreseen. In the following section, we will refer to $t_{\scalebox{.7}{obs}}=10$ yrs for both interferometers.\footnote{Note that, given the monochromatic nature of our targets, results are essentially unaffected if the data are obtained in a continuous stream, rather than collected along a longer mission with a duty cycle of less than 100\%.}
Fig.\,\ref{cfr_sens} shows the sensitivities of LISA and $\mu$ARES in the frequency range relevant to our study.

\section{results}
In order to compute the GW signal arising from the population of PBHs described in the previous sections, we run a series of Monte-Carlo simulations randomly sampling the underlying distribution with 4000 PBHs of 1\,M$_{\odot}$. From the sampled population we then compute the resulting GW signal.  We explicitly make a distinction between resolved events and unresolved ones, the latter combining to build-up a stochastic background. Our final results are then obtained by averaging the GW signal over a statistically significant number of simulations. 
In the following, we analyse the two different types of signals (resolved and background). All relevant figures are reported in Table~\ref{tab:cases}.

\subsection{\label{subsec_res_BG}Resolved events}
In estimating the distribution of resolved sources, we deem an event ``resolvable" whenever the two following criteria are simultaneously satisfied:
\begin{itemize}
\item the event has SNR\,$>8$;\footnote{Circular EMRIs like the ones considered here are essentially monochromatic sources featuring a waveform very similar to that of galactic white dwarf binaries, for which resolvability down to SNR\,$\approx8$ has been demonstrated in early LISA mock data challenges \cite{2010PhRvD..81f3008B}.}
\item no more than 1 event falls within a given frequency resolution bin\footnote{The frequency resolution of the data is defined as the inverse of $t_{\scalebox{.7}{obs}}$, so that for $t_{\scalebox{.7}{obs}}=10\,(4.5)$ yrs, the corresponding frequency resolution will be $3\,(7)\times10^{-9}$\,Hz.} \cite{sesa08}.
\end{itemize}

\begin{figure}
\begin{subfigure}{.5\textwidth}
{\centerline{\includegraphics[scale=0.550]{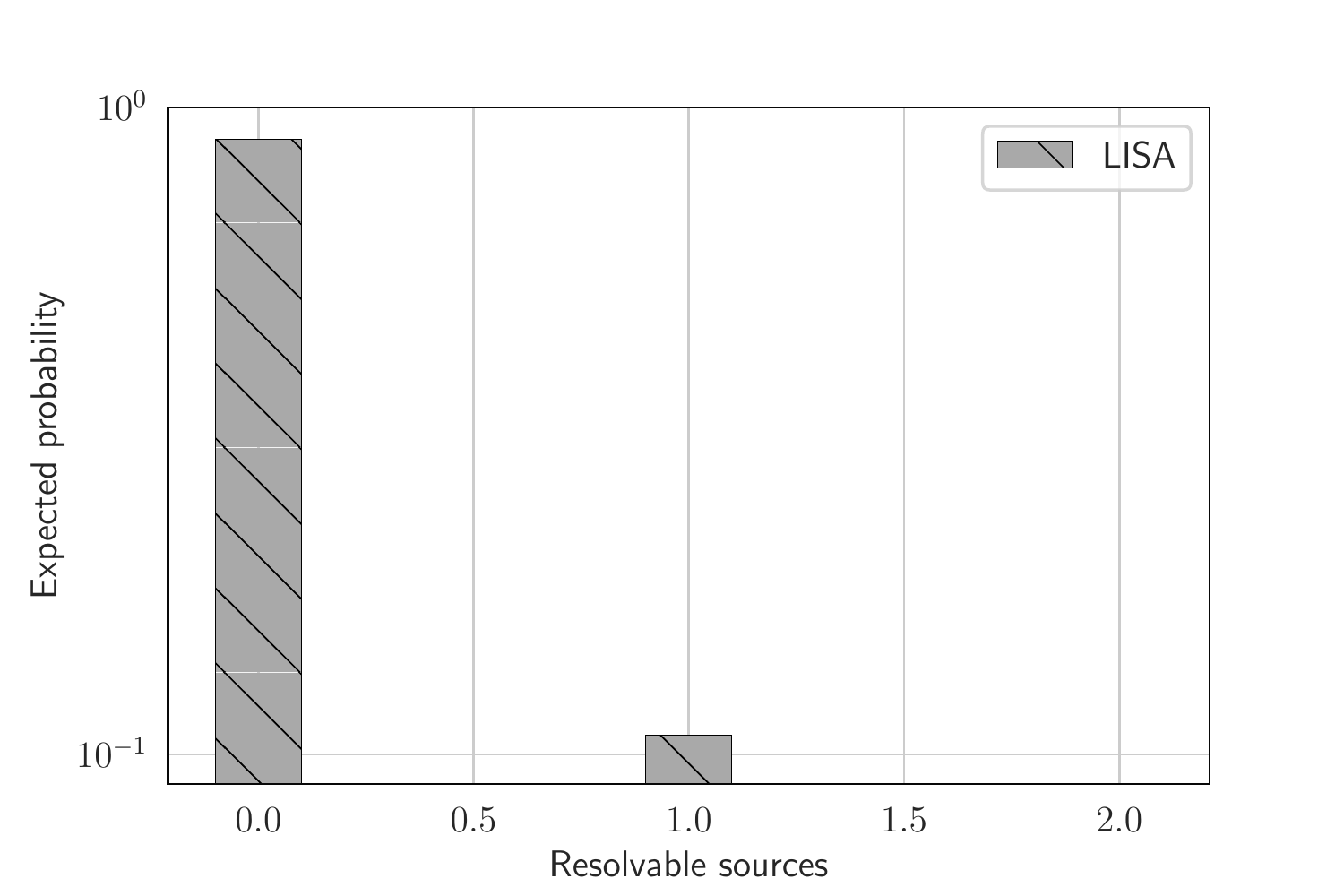}} }
\vspace{-0.1cm}\caption{}
\label{quante_risolte_LISA}
\end{subfigure}\quad\hfill
\begin{subfigure}{.5\textwidth}
{\centerline{\includegraphics[scale=0.55]{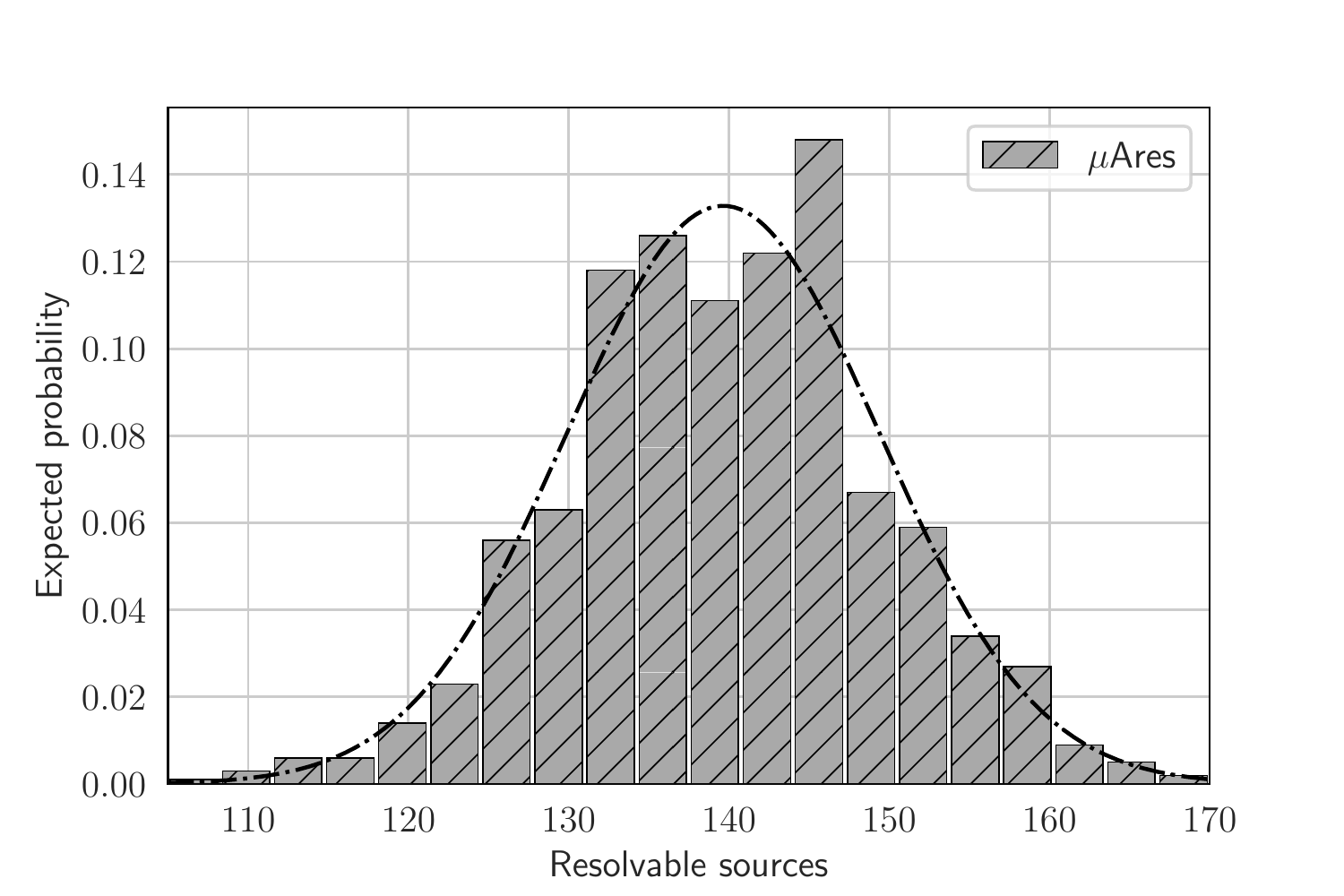}} }
\vspace{-0.1cm}\caption{}
\label{quante_risolte_Ares}
\end{subfigure}
\captionsetup{format=plain}
\vspace{-.35cm}\caption{\raggedright Expected probability of finding a given number of PBHs, for LISA (a) and $\mu$Ares (b). The latter distribution is fitted with a Gaussian with parameters mean $=139.6$ and standard deviation $=9.7$.}
\label{quante_risolte}
\end{figure}

As a first step, through Monte-Carlo simulations we randomly select 4000 PBHs (so as to cope with the mass constraints given by \cite{gravity21}) from the underlying distribution, constructing a catalog of potential sources. Then, for each source in the catalog, we compute the GW signal and its SNR according to the LISA and $\mu$ARES sensitivities. 
As in a typical catalog realization LISA would resolve from zero to a maximum of 1 event, in order to have a statistically significant figure we run a total of 1,000 simulations. 
Fig.~\ref{quante_risolte} shows the probability, computed over 1,000 Monte-Carlo realizations, that a catalog contains a given number of resolved events. Regarding LISA, it is apparent how the vast majority of realizations contain no detection whatsoever, with few simulations resulting in just 1 event. Statistically, there is a $\simeq 11$\% probability that one PBH might be resolved by LISA in 10 year of data collection, with the remaining 89\% probability of null detection\footnote{Detection probabilities are only slightly modified by a change in $t_{\scalebox{.7}{obs}}$, which in turn affects the frequency bin width and the number of cycles in the characteristic strain. For instance, halving $t_{\scalebox{.7}{obs}}$ would reduce the SNR by a factor $\sqrt{2}$.}.

\begin{figure}
\begin{subfigure}{.5\textwidth}
{\centerline{\includegraphics[scale=0.54]{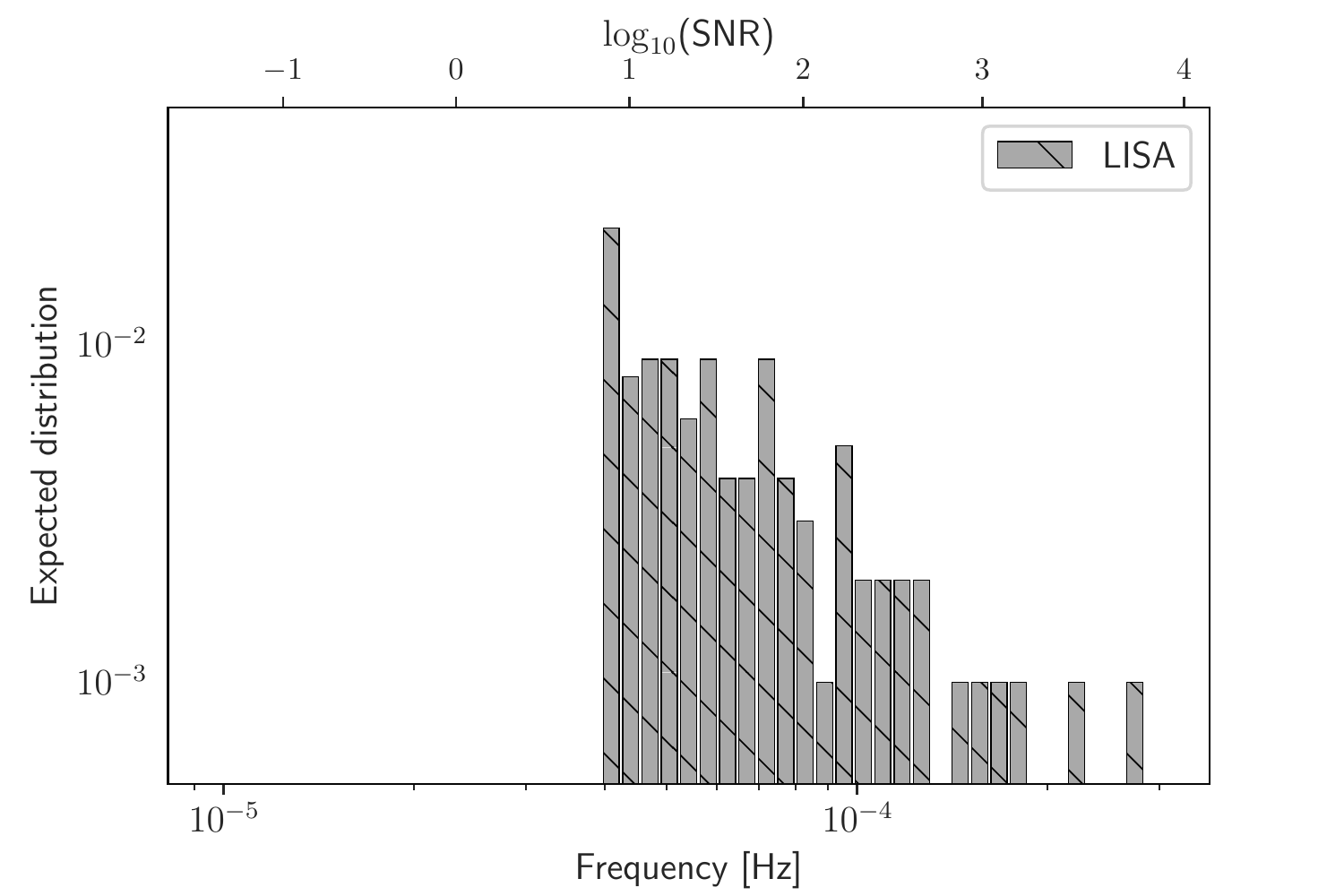}} }
\vspace{-0.1cm}\subcaption{}
\label{freqresLISA1000}
\end{subfigure}\quad\hfill
\begin{subfigure}{.5\textwidth}
{\centerline{\includegraphics[scale=0.54]{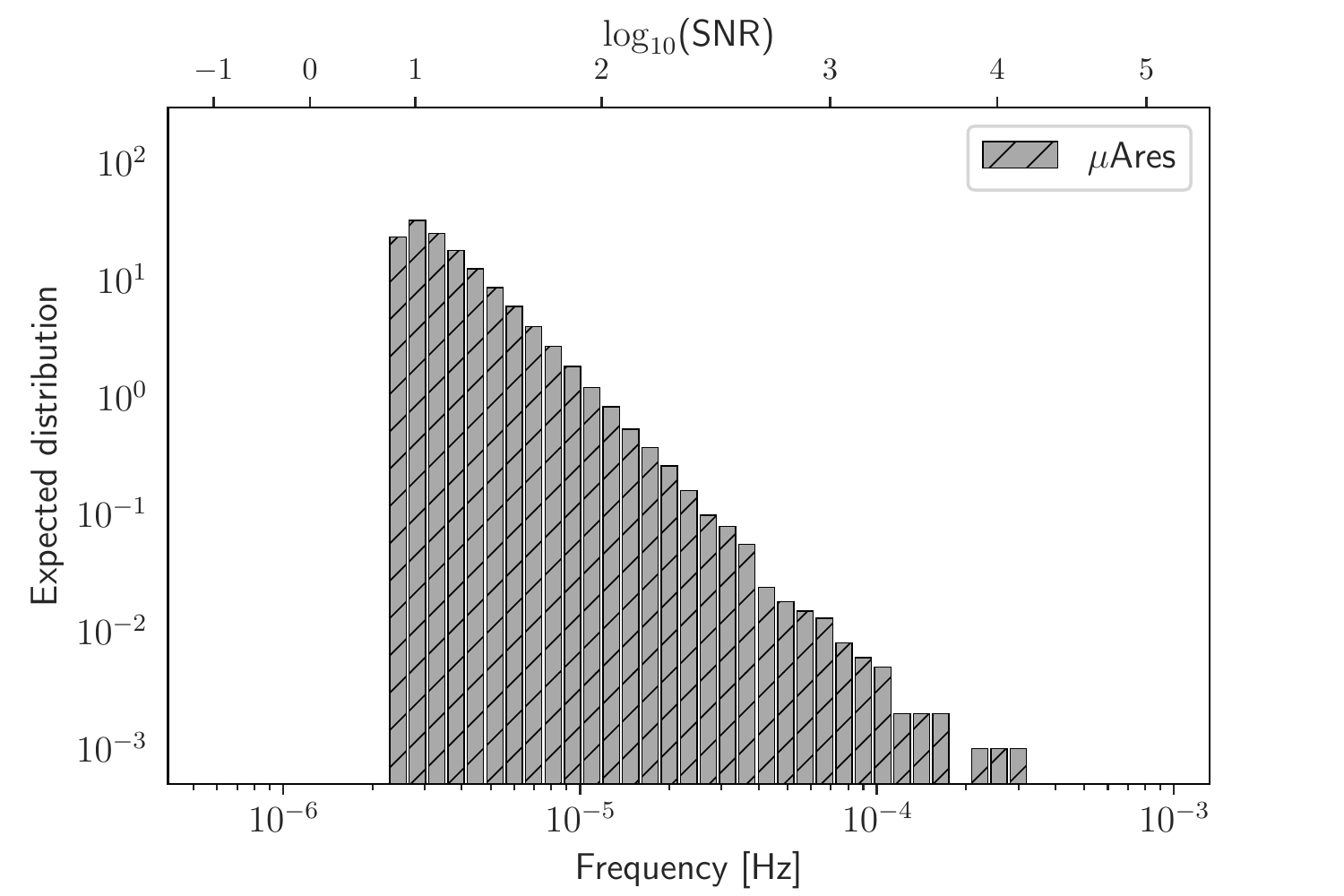}}}
\vspace{-0.1cm}\subcaption{}
\label{freqresAres1000}
\end{subfigure}
\vspace{-.35cm}\caption{\raggedright Distribution of expected sources resolvable by LISA (a) and $\mu$Ares (b), over the expected instrumental lifetime, as function of frequency and SNR.}
\label{freqres1000}
\end{figure}

Given its much higher sensitivity at low frequencies, the outcome for $\mu$Ares is strikingly different, resulting in an average of $\simeq 140$ detected PBHs in 10 years. The probability distribution, again computed over 1,000 realizations, is well fit by a Gaussian distribution with mean and standard deviation of $139.6$ and $9.7$, respectively. Under the assumption that the unresolved matter within the Galactic center is entirely formed by PBHs, this means that $\mu$Ares would have a chance of $99.9$\% of resolving a minimum of 110 solar-mass PBHs orbiting \sgrA. 
Under our assumption of circular orbits and inclination-and-polarization averaged strain, there exist 1:1:1 relations among the radial distance of a PBH to \sgrA, its GW frequency and the SNR of the event. From the source catalogs we can then compute frequency and SNR distributions, shown in Fig.~\ref{freqres1000}. The distributions are obtained adopting a logarithmic binning in frequency, and are normalized so that the sum of the histogram heights gives the average number of resolved events, i.e., 0.11 in the case of LISA and 140 for $\mu$Ares. Regarding LISA, all we can say is that the single one event possibly resolved during 10 years of data collection would have a higher chance to fall in the frequency range $3.5 \times 10^{-5}\lesssim f \lesssim 6 \times 10^{-5}$ Hz, corresponding to $8 \lesssim {\rm SNR} \lesssim 30$. $\mu$Ares, instead, will produce a genuine distribution of resolved events, covering at least one decade in frequency with an SNR as large as few hundreds.

Finally, we note that the detection statistics is mainly driven by the SNR\,$>8$ constraint, as for $f\gtrsim 10^{-5}$\,Hz there are, on average, less than 1 PBH per frequency bin anyway. This applies to both interferometers, 
although in the case of $\mu$Ares some overlap may occur in the lower frequency bins, where the number of resolvable sources is larger (Fig.~\ref{freqres1000}, panel b).

The figures reported here, and in next \S~\ref{subsec_intro} as well, would be only marginally affected by a different choice of the SNR threshold. As an example, an SNR threshold of 5 would rather produce a $\lesssim 1\%$ increase in the chances of detection by LISA and a $\simeq 5\%$ increase in the number of sources resolvable by $\mu$Ares.

\subsection{\label{subsubsec_BG}Stochastic background}

Many PBHs, if not almost all as in the case of LISA, do not satisfy the criteria for being resolvable. Still , their cumulative GW signal could well produce a background signal whose SNR would be above detection threshold.  

In order to estimate the amplitude of the background signal, we use Eq.\,(7) of \cite{sesa16},
\begin{equation}
(S/N)^{2}_{\scalebox{.7}{bkg}} = t_{\scalebox{.7}{obs}}\int\gamma(f)\frac{h_{c,\scalebox{.7}{bkg}}^{4}(f)}{4f^{2}S_n(f)^{2}} df,
\label{eq:SNRbkg}
\end{equation} 
where again we made use of the position and polarization-averaged sensitivity $S_n(f)$. 
According to Fig.\,4 in \cite{thraneromano} and to \cite{sesa16}, the so-called response function is $\gamma(f)\approx1$ in the relevant frequency range, while for the characteristic strain $h_{c,\scalebox{.7}{bkg}}^{2}$ we use (from \cite{2017arXiv170200786A})
\begin{equation}
h_{c,\scalebox{.7}{bkg}}^{2} =\mathlarger{\sum}_{i} \frac{h_{i}(f)^{2}f_{i}}{\Delta\,f}\equiv  \mathlarger{\sum}_{i} h_{i}(f)^{2}\mathcal{N}_{\scalebox{.7}{cyc}}(f).
\label{eq:cstrain}
\end{equation}
In the above Eq.\,\ref{eq:cstrain} the summation is intended over the whole catalog {\it excluding all resolved  sources}.
As discussed in \cite{subtraction}, this is a somewhat optimistic approach, as it implicitly assumes a free-of-errors estimate of source parameters, and a optimal waveform subtraction. The monochromatic nature of the PBHs considered here makes this approach reliable.

\begin{figure}
\begin{subfigure}{.5\textwidth}
{{\includegraphics[scale=0.55]{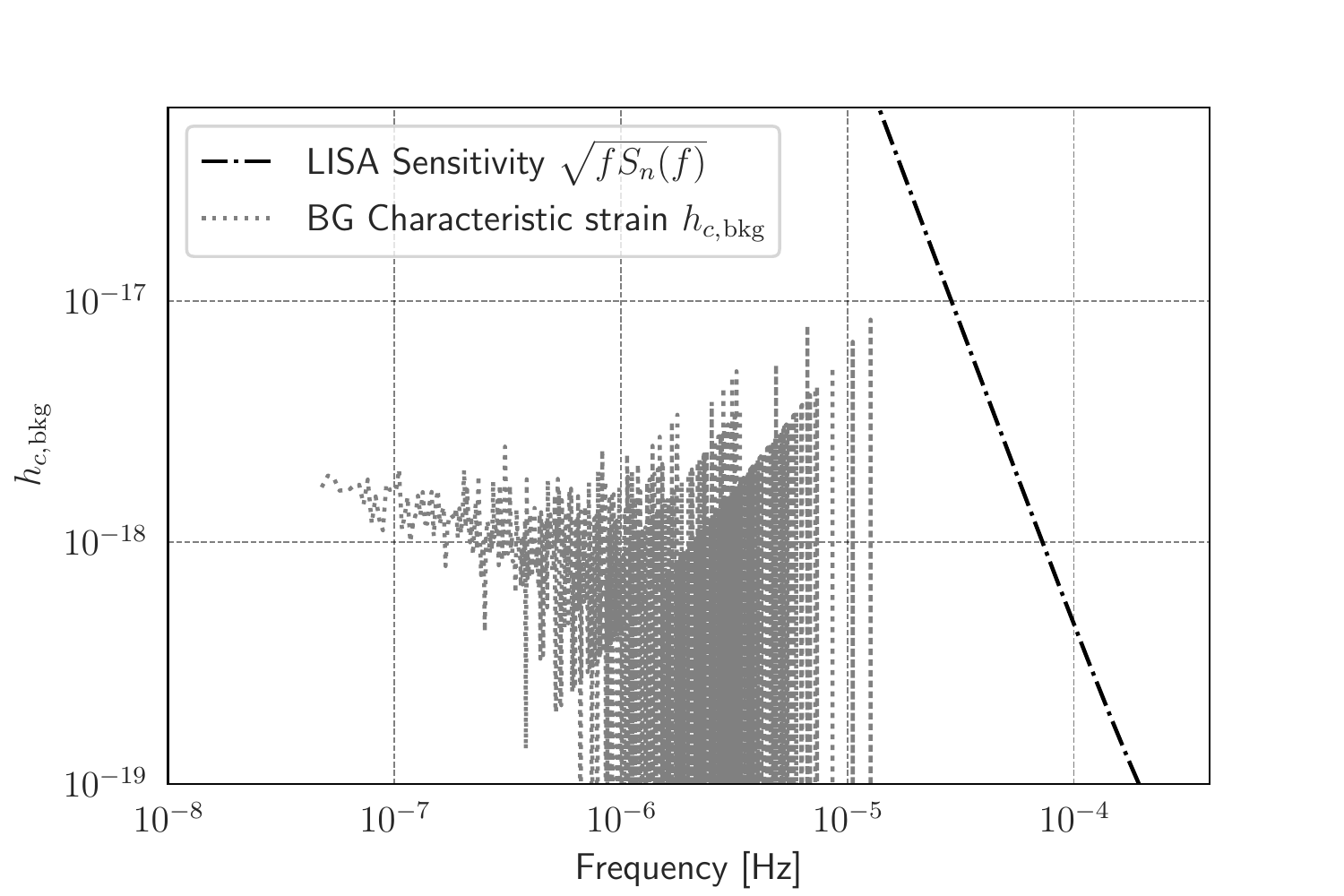}} }
\caption{}
\label{plotLISA}
\end{subfigure}
\quad\hfill
\begin{subfigure}{.5\textwidth}
{{\includegraphics[scale=0.55]{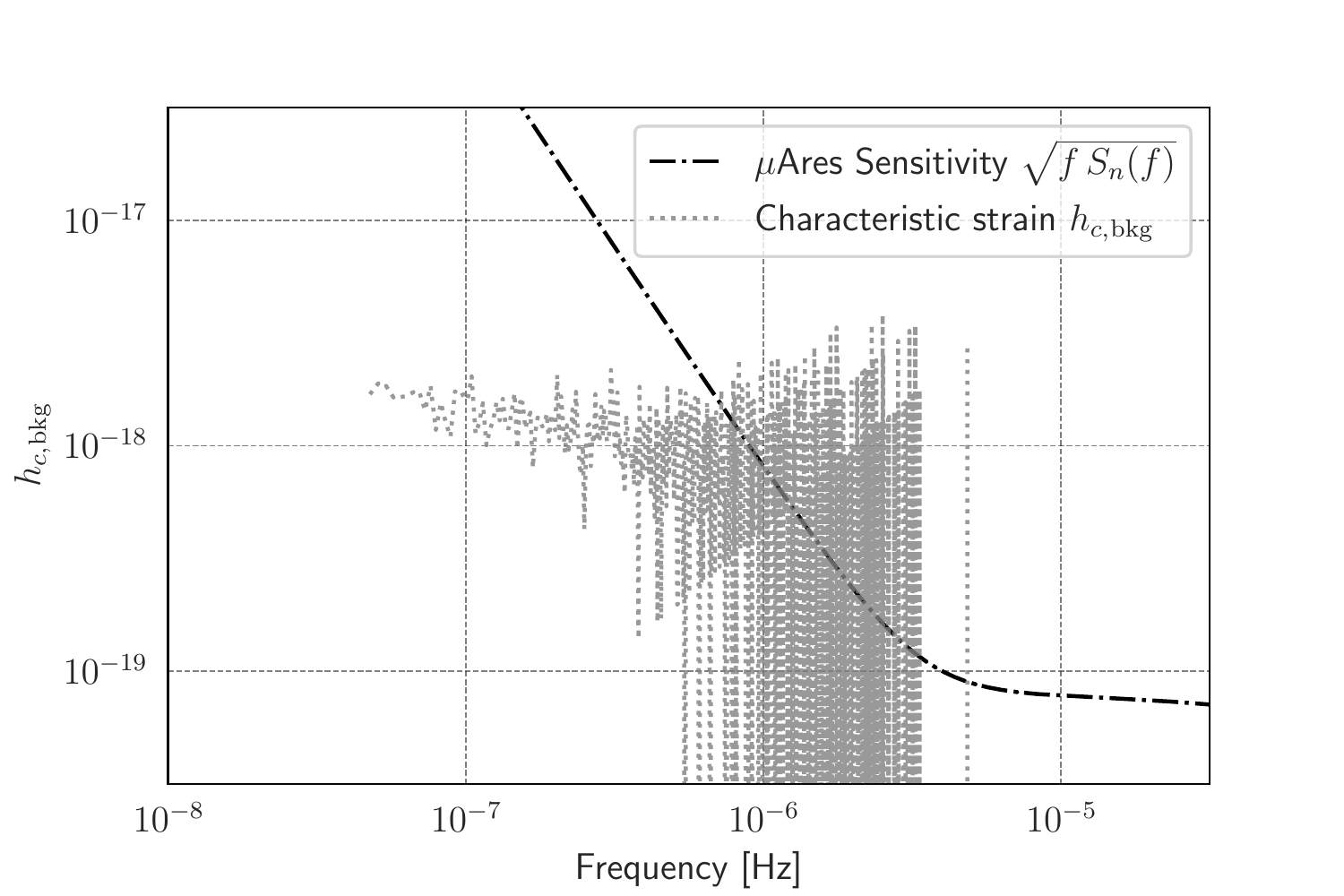}} }
\caption{}
\label{hcAres}
\end{subfigure}
\vspace{-.35cm}\caption{Characteristic strain binned and summed in frequency bins, plotted against the LISA and $\mu$Ares sensitivity curves.}
\label{curveAres}
\end{figure}

\begin{figure}
\begin{subfigure}{.5\textwidth}
{\centerline{\includegraphics[scale=0.55]{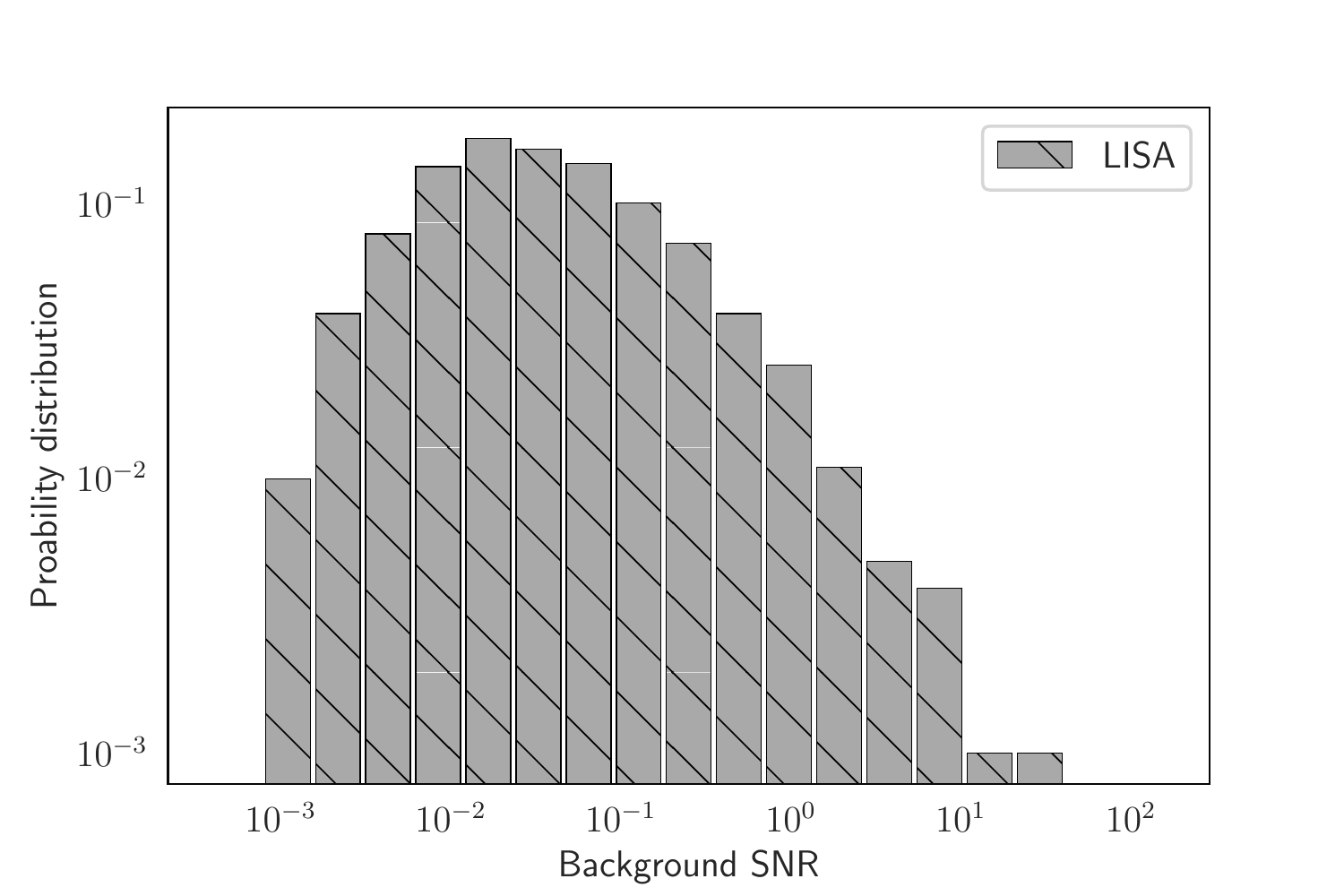}} }
\caption{}
\label{BG_L}
\end{subfigure}\quad\hfill
\begin{subfigure}{.5\textwidth}
{\centerline{\includegraphics[scale=0.55]{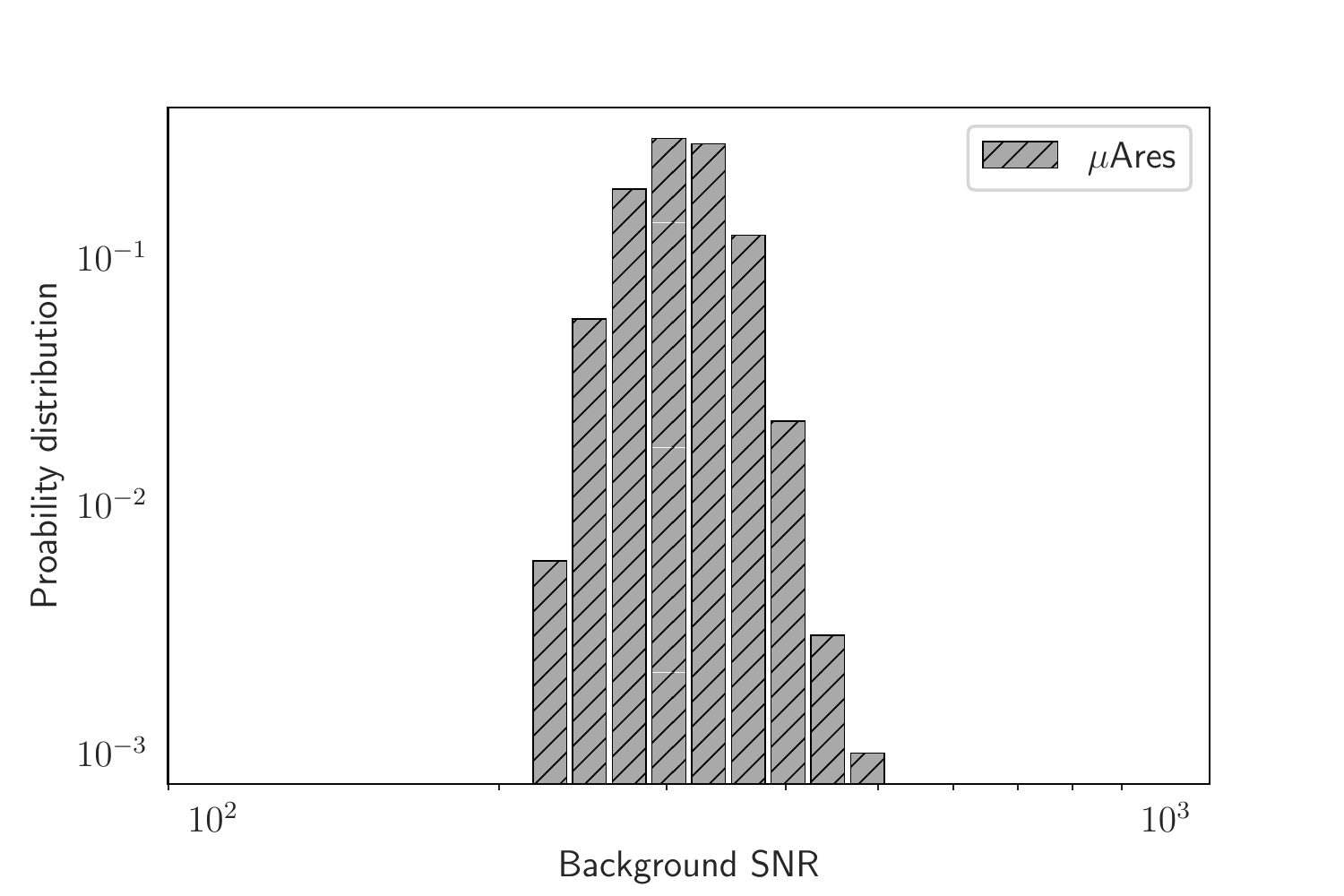}} }
\caption{}
\label{BG_A}
\end{subfigure}
\vspace{-.35cm}\caption{\raggedright Expected distribution of the GW background SNRs for LISA (a) and $\mu$Ares (b).}
\end{figure}

In Fig.\,\ref{plotLISA} and Fig.\,\ref{hcAres} we show $h_{c,\scalebox{.7}{bkg}}(f)$ for all unresolved sources in a 10-year-long stream of LISA and $\mu$Ares data, respectively. While at the lowest frequencies the characteristic strain resembles that of a typical background noise, at higher frequencies the relatively low number of sources gives the signal a ``pop-corn" flavour, with frequency bins filled by more than one source interloped by empty ones. It is interesting to note how in the case of LISA, given its much lower sensitivity, many high frequency sources are counted in the non-resolved pool, and hence do contribute to the background. For $\mu$Ares, instead, sources at high frequencies (and hence high strains) will be always resolvable, consequently the background will not extend in the $10^{-5}$ Hz regime. 
The effect is apparent also when we plot the expected probability density distribution of the background SNR (Eq.\,\ref{eq:SNRbkg}), shown in Fig.\,\ref{BG_L} in the case of LISA, and in Fig.\,\ref{BG_A} in the case of $\mu$Ares. The results are again obtained averaging over 1,000 simulation runs, and refer to a 10-year-long data stream. The stochastic background in the case of LISA spans a quite large range in SNR, but it still has a very low chance to be detectable during the mission lifetime. On the contrary, in the case of $\mu$Ares the whole of unresolved sources combine to produce a GW background which would be observable with a SNR of few hundreds.

\subsection{\label{subsec_intro}Dependence upon PBH mass and density distribution}

In this section we relax our assumptions regarding the typical mass of PBHs, and the details of the mass density profile, vetting how our results  depend upon the specific choice of input parameters.  
Regarding possible PBH masses connected to GW detection, we can limit the pertaining parameter space as follows. Based on various current upper limits (see, e.g., \cite{carr2021}), a window exists between $10^{-16}$ and $10^{-11}$\,M$_{\odot}$ where 100\% of dark matter could be in the form of PBHs without violating any observational constraint. However, such PBHs would be much too light to enter the GW-dominated regime\footnote{Note that if such light PBHs do happen to exist, they would be largely evacuated from the Galactic center because of mass segregation.}. This is true up until $\sim 10^{-4}$\,M$_{\odot}$, where, still according to \cite{carr2021}, the maximum allowed dark matter fraction in the form of PBHs happens to be $\approx 1\%$. A more operational mass limit for PBHs can be set by considering the minimum mass orbiting \sgrA ~at the ISCO (i.e.\ at 1.2\,$\times10^{-6}$\,pc) whose GW losses are above the detection threshold. Such mass turns out to be $\gtrsim 5\times10^{-5}$\,M$_{\odot}$ for the LISA sensitivity. Note that, if $\mpbh \lesssim 0.01$\,M$_{\odot}$, the time to coalescence at the peak of the density distribution (i.e., where most of the PBHs would be) exceeds the Hubble time. Having  considered all this, we redo our analysis allowing the PBHs to have masses as low as $0.01$\,M$_\odot$, for which the maximum possible dark matter fraction in PBHs is between a few and 10\%,  and as large as 10\,M$_{\odot}$, for which similar constraints exist \cite{carr2021}. Note that the fixed total mass allowed in our models sets the number of PBHs orbiting \sgrA ~to 4000 M$_\odot/\mpbh$. 
We also change the PBH density distribution $\rho(r)$, testing two alternative models, different from the Spiked NFW profile adopted so far: an isothermal profile, $\rho(r)\propto r^{-2}$, and a Bahcall-Wolf profile, $\rho(r)\propto r^{-7/4}$ \cite{bahcallwolf}, again normalized to 4000 M$_\odot$ within the S2 pericenter, and again shaped in the inner region by GW losses. The resulting distributions are shown in Fig.\,\ref{rhoGW2br}. It is important to notice that while for the Spiked NFW case the peak of the number distribution is at the sink radius $r_{\textrm{sink } }$ (i.e., the distance form \sgrA ~where two-body relaxation and GW timescales are equal), in the case of an isothermal profile we have the same number of PBHs at every $r>r_{\textrm{sink } }$, and that for a Bahcall-Wolf profile the higher PBH density occurs at the largest allowed distance. This very fact bears important consequences when forecasting detection figures. Results of our analysis are reported in Table~\ref{tab:cases}, where all the tested cases are summarized. 
Generally speaking, when we change $\mpbh$, we are dealing with two competing effects: on the one hand lighter PBHs are more numerous, on the other hand the GW signal from a single source is weaker. On top of that, the GW signal has to be folded into the sensitivity curve of the interferometer under consideration. It is the interplay among these three effects that sets the outcome of the experiment that we performed. Having this in mind, it is more practical discussing results for LISA and $\mu$Ares separately. In the case of LISA, it is apparent that the larger the number of PBHs (i.e., the lower $\mpbh$), the larger the number of resolved events, in agreement with naive expectations. Indeed, the low frequency sensitivity of LISA scales approximately as $f^{-2.5}$ (see Fig.\,\ref{cfr_sens}) and, according to \cite{sesa20}, this would produce a number of resolved events $N_{\textrm{res}} \propto \mathcal{M}_c^{-5/11}$.
As $\mpbh\ll M_{\lw{BH}}$ implies $\mathcal{M}_c\propto \mpbh^{3/5}$, then $N_{\textrm{res}}\propto \mpbh^{-3/11}$.
An increase of 3 dex in $\mpbh$ would then result in a decrease of resolved events of a factor $\approx 7$, in line with our findings. Note however that, for $t_{\scalebox{.7}{obs}}=10$ yrs, we find $N_{\textrm{res}}<1$, i.e., we can merely interpret such number as a probability of detecting a single resolved event during the mission. Such probability is as large as 60\% in the case of $\mpbh=0.01$\,M$_\odot$ with an isothermal profile. Note also that while the Spiked NFW and isothermal cases give comparable results, the shallow Bahcall-Wolf distribution reduces the probability of detection by an order of magnitude. Rather interestingly, the median SNR of the possible detection is very similar in all tested cases. 

Concerning the possibility of a LISA detection of background noise arising from the population of PBHs, though the SNR increases for lower masses, none of our models predict a statistically significant stochastic signal in the LISA data stream.\\
For $\mu$Ares results are somewhat less straightforward. In this case (see Fig.\,\ref{cfr_sens}) the sensitivity curve 
features a sort of plateau for $f\simgt 10^{-5.5}$~Hz, basically shaped by the WD stochastic background. For large values of $\mpbh$, $N_{\textrm{res}}$ is determined by the steep slope of the sensitivity curve for $f\lesssim 10^{-5.5}$\,Hz. This is because in this case the GW signal is strong enough that PBHs enter the observability band well below $10^{-5.5}$\,Hz. Reducing the typical PBH mass would then increase $N_{\textrm{res}}$, because of the very same reason seen in the case of LISA. However this is true only down to a certain mass, for which PBHs start entering the observability band above $10^{-5.5}$\,Hz, i.e., in the flat part of the sensitivity curve. Further lowering the mass would lead more and more sources to fall well below detectability threshold. The combined effects of PBH typical mass {\it and} sensitivity then creates a ``sweet spot" for $N_{\textrm{res}}$ happening to be just around $\mpbh\simeq 1$\,M$_\odot$, as reported in  Table~\ref{tab:cases}. As in the case of LISA, the shallow Bahcall-Wolf profile gives an order of magnitude less detections. Still, even in this unfavorable circumstance, only for the lowest PBH mass the number of resolved events in 10 years is below unity. Finally, regarding background detectability, in all cases $\mu$Ares would produce a signal with a very high SNR.

\section{\label{sec_concl}discussion and conclusions}

If a distribution of PBHs is present in our Galaxy, it might concentrate at the Galactic center, where 
these objects would be expected to orbit around the central 
massive black hole \sgrA, thus constituting possible sources for gravitational wave detectors. Assuming a stationary 
distribution of PBHs subject to two-body relaxation and gravitational-wave driven infall toward \sgrA, and complying with the mass limits posed by S2 pericenter precession~\cite{gravity21},
we have computed the expected (resolved and unresolved) GW signal detectable by future space-borne observatories such as LISA and $\mu$Ares.
Although simplified, our model shows that there is a $\simeq10\%$ chance for LISA to resolve a 1\,M$_{\odot}$ primordial black hole during a 10 years observation time, while even less likely is the detection of a background signal. A solid chance of detection might instead be expected from the proposed space-borne interferometer $\mu$Ares, whose higher sensitivity would allow one to resolve from several to more than one hundred PBHs, regardless of the actual typical mass or density profile, and to detect an unresolved background with a signal to noise ratio well above detection threshold.

\begin{table*}

\begin{tabular}{y{1.6cm}x{1.8cm}x{1.6cm}y{1.7cm}x{1.6cm}x{1.6cm}y{1.6cm}x{1.6cm}x{1.6cm}x{1.6cm}}
\toprule
\mytab{\vspace{0.05cm}{PBH Mass} \newline {[M$_{\odot}$]}} &
\mytab{\vspace{0.05cm}{Sink radius} \newline {[$\times10^{-5}$\,pc]}} &
\mytab{\vspace{-0.1cm} {Sink}\newline{frequency}\newline{ [$\times10^{-6}$Hz]}}&
\mytab{\vspace{0.05cm}{Time to\newline merger [yr]}}&

\multicolumn{3}{c@{}}{LISA}&
\multicolumn{3}{c@{}}{$\mu$Ares}\\
 \cmidrule(lr){5-7} \cmidrule(lr){8-10}\\
 &&&&
 \mytab{\vspace{-0.61cm}{Resolved\newline PBHs}}&
 \mytab{\vspace{-0.43cm} SNR$_{\scalebox{.80}{Res}}$}&
 \mytab{\vspace{-0.43cm} SNR$_{\scalebox{.80}{BG}}$}&
 \mytab{\vspace{-0.61cm} {Resolved\newline PBHs}}&
 \mytab{\vspace{-0.43cm} SNR$_{\scalebox{.80}{Res}}$}&
 \mytab{\vspace{-0.43cm} SNR$_{\scalebox{.80}{BG}}$}\\ 
\mytab{(1)}&\mytab{(2)}&\mytab{(3)}&\mytab{(4)}&\mytab{(5)}&\mytab{(6)}&\mytab{(7)}&\mytab{(8)}&\mytab{(9)}&\mytab{(10)}\\

\midrule
 
         10 \footnotemark[1] & ${6.40}$ &2.75& ${5.15\times10^{7}}$ & 0.05 & 22.5\comm{$\pm...$} & 0.004 & 46 & 44.1 & 146 \\
         5 \footnotemark[1] & ${6.29}$ & 2.82& ${9.66\times10^{7}}$ & 0.049 & 20.5\comm{$\pm...$} & 0.007 & 70 & 33.8 & 1869 \\
         3 \footnotemark[1] & ${6.22}$ & 2.87& ${1.54\times10^{8}}$ & 0.08 & 25.4\comm{$\pm...$} & 0.011 & 92 & 27.0 & 2970 \\
         1 \footnotemark[1] & ${6.08}$ & 2.97& ${4.21\times10^{8}}$ & 0.11 & 28.4\comm{$\pm...$} & 0.03 & 140 & 17.1 & 304 \\
         0.1\footnotemark[1] & 5.89 &3.11& $3.54\times10^9$ & 0.26 & 20.3 & 0.2 & 53\comm{$\pm7$} & 11.4 & 329 \\
         0.01\footnotemark[1] & $5.61$ & 3.35& ${3.50\times10^{10}}$ & 0.42 & 16.8\comm{$\pm...$} & 1.2 & 4 & 12.9\comm{$\pm...$} & 863 \\

\cmidrule[.01pt](){1-10}
        10\footnotemark[2] & 4.27 &5.04& ${1.02\times10^7}$ & 0.08& 22.1\comm{$\pm...$} & 0.01\comm{$\pm2$} & 32 & 63.7\comm{$\pm...$} & 227\comm{$\pm...$}  \\
        5\footnotemark[2] & 4.20 & 5.17& ${1.92\times10^7}$ & 0.12 & 24.0\comm{$\pm...$} & 0.03\comm{$\pm2$} & 52 & 50.0\comm{$\pm...$} & 480\comm{$\pm...$}  \\
        3\footnotemark[2] & 4.16 & 5.25& ${3.07\times10^7}$ & 0.13 & 22.8\comm{$\pm...$} & 0.05\comm{$\pm2$} & 72 & 40.7\comm{$\pm...$} & 984\comm{$\pm...$}  \\
        1\footnotemark[2] & 4.07 & 5.42& ${8.45\times10^7}$ & 0.16 & 21.4\comm{$\pm...$} & 0.11\comm{$\pm2$} & 135 & 24.4\comm{$\pm...$} & 3017\comm{$\pm...$}  \\
         0.1\footnotemark[2]& 3.91 & 5.76& ${7.20\times10^8}$ & 0.32 & 20.7\comm{$\pm...$} & 0.88\comm{$\pm...$} & 91 & 11.2\comm{$\pm...$} & 7037\comm{$\pm$}  \\
         0.01\footnotemark[2] & 3.77 &6.08& ${6.25\times10^9}$ & 0.61 & 21.8\comm{$\pm...$} & 4.13\comm{$\pm$} & 8 & 12.5\comm{$\pm...$} & 2152 \\
         
\cmidrule[.01pt](){1-10}
         10\footnotemark[3] & 9.56 & 1.51& ${2.57\times10^8}$ & 0.008 & 25.9\comm{$\pm...$} & $7.85\times10^{-6}$\comm{$\pm$} & 11 & 22.8\comm{$\pm...$} & 143\comm{$\pm...$}  \\
        5\footnotemark[3] & 9.43 & 1.54& ${4.86\times10^8}$ & 0.006 & 29.3\comm{$\pm...$} & $1.7\times10^{-5}$\comm{$\pm2$} & 14 & 19.1\comm{$\pm...$} & 254\comm{$\pm...$}  \\
        3\footnotemark[3] & 9.33 & 1.56& ${7.78\times10^8}$ & 0.014 & 16.9\comm{$\pm...$} & $2.57\times10^{-5}$\comm{$\pm$} & 16 & 18.0\comm{$\pm...$} & 343\comm{$\pm...$}  \\
         1\footnotemark[3]& ${9.15}$ & 1.61& ${2.16\times{10^9}}$ & 0.01 &  21.1\comm{$\pm...$} & $7.25\times10^{-5}$\comm{$\pm...$} & 18 & 15.7\comm{$\pm...$} & 515\comm{$\pm...$}  \\
         0.1\footnotemark[3]& ${8.81}$ & 1.70& ${1.86\times10^{10}}$ & 0.02 & 17.3\comm{$\pm...$} & $5.14\times10^{-4}$\comm{$\pm0.2$} & 5 & 12.1\comm{$\pm...$} & 594 \\
         0.01\footnotemark[3]& ${8.53}$ &1.79&  ${1.63\times10^{11}}$ & 0.04 & 20.7\comm{$\pm...$} & 0.003\comm{$\pm0.4$} & 0.1& 14.0\comm{$\pm...$} & 137  \\
 \bottomrule
\end{tabular}
\label{tab:table3}
\caption{\label{tab:cases} Results for different PBH masses and density profiles (Spiked NFW\,($^{\text{a}}$), isothermal sphere\,($^{\text{b}}$), Bahcall-Wolf\,($^{\text{c}}$)). (1) PBH mass; (2) sink radius, i.e.\ the distance from \sgrA \,where the crossing between the two body relaxation and GW regimes occurs; (3) corresponding GW frequency; (4) corresponding time to coalescence by gravitational waves infall; (5) average value of detection probability of at least 1 PBH by LISA; (6) median value of the corresponding SNRs; (7) median value of the background SNR for LISA; (8) average value of detectable PBHs by $\mu$Ares; (9) median of corresponding SNRs; (10) median of the background SNR for $\mu$Ares.}
\end{table*}

Another important question has to do with inferring the nature of the detected sources. In other words, should LISA detect a 1\,M$_{\odot}$ source orbiting \sgrA, will it be possible to 
distinguish between a primordial black hole  from 
an astrophysical object, such as a star, or a brown dwarf?
First, we should notice that measuring the (chirp) mass of a source
(and thus the mass of the PBH)
is only possible if the frequency is evolving, i.e., if the source is not 
completely monochromatic.
From the quadrupole formula, the source frequency's rate of change is~\cite{Peters:1964zz}
\begin{equation}
    \dot{f}=\frac{96}{5}\pi^{8/3}\Bigl(\frac{G \mathcal{M}_{c}}{c^3}\Bigr)^{5/3}f^{11/3}\,.
\end{equation}
The frequency resolution $\Delta f$ of an experiment is 
 the inverse of $t_{\scalebox{.7}{obs}}$, 
 i.e., $\Delta f\simeq 3\times 10^{-9}$ Hz for 10 years of
 observation. At the highest resolved frequency by LISA (see Fig.\,\ref{freqresLISA1000}), i.e.\ $f\simeq3\times10^{-4}$\,Hz, the time
 needed for the frequency to change by $\Delta f$ is 19 days.
At the lowest resolved frequency by LISA (again see Fig.\,\ref{freqresLISA1000}), $f\simeq4\times10^{-5}$\,Hz, the time goes up to 87 years. The cutoff frequency, where it takes exactly 10 years 
for the frequency to change by $\Delta f$, is $7.2\times10^{-5}$\,Hz. Should the data stream span less than 10 years, say 4.5 years, such cutoff frequency would rise to $\simeq1.1\times10^{-4}$\,Hz, taking into account that $\Delta f$ would change accordingly. 
Therefore, we do {\it not} expect  the mass of the PBH candidate to be measurable for all the detected events.
Even for the events with measurable mass, a possible astrophysical origin should be considered. To distinguish between a star and a PBH, one may consider tidal effects. In more detail, a star would be tidally disrupted at the tidal disruption radius \cite{evogalaxy} $r_t=R_{\star}(\frac{M_{BH}}{m_{\star}})^{1/3}$ with $R_{\star}$ and $m_{\star}$ the star's radius and mass. For a solar-type star the tidal radius is $r_{t,\odot}\approx3.7\times10^{-6}$\,pc, or 9 Schwarzschild radii.
As can be easily understood by 
comparing e.g.\ to our Fig.\,\ref{rhoGW2br}, this is way too close to \sgrA ~to prove useful for telling 1\,M$_{\odot}$ PBHs and stars apart. For comparison, at a distance of $10^{-5}$\,pc, which is well within the range of  Fig.\,\ref{rhoGW2br}, 
the typical stars that would be tidally disrupted would be ones like S2 (i.e.\ a B0 star with mass $\simeq14\,M_{\odot}$ and radius $\simeq7\,R_{\odot}$). 
Clearly, tidal effects can be more subtle, as a deformed star, even if not disrupted, would show characteristic tidal effects in the gravitational waveforms. However, those would only be observable if the source frequency is evolving  (cf.~the discussion on the mass estimate right above).
Even more difficult would be to distinguish a neutron star from a PBH of similar mass; only in the case of a pulsar would such distinction be easily performed.

These preliminary considerations show that, although the detection of PBHs in the GC might be feasible in the future, recognizing their PBH nature might not be at all straightforward. In future work, we will therefore study in more detail the parameter
estimation capabilities of LISA and $\mu$Ares, focusing on the distinctness between PBHs and stars/brown dwarfs.
Additionally, we will also assess the impact of eccentric PBH orbits on our results. In fact, a relaxed isotropic cusp of PBHs in the Galactic center is expected to feature a thermal eccentricity distribution, i.e.\ $p(e)\propto e$. Therefore, despite GW-driven circularization, we expect the overall signal to be dominated by eccentric sources. The importance of eccentricity is twofold. On the one hand, eccentric sources emit at higher frequencies, which might significantly increase the chances of LISA to see such systems. On the other hand, eccentric sources evolve much more rapidly, thus allowing for a better determination of the source mass, therefore helping the assessment of the source nature.

In closing, we underline that, compared to LISA and $\mu$Ares, thanks to their sensitivity at higher frequencies ground-based interferometers such as the next Einstein Telescope \cite{Punturo2010} will play a complementary role in the search for GWs emitted by PBHs, e.g., in the detection of binaries of such objects \cite{PBHB1, PBHB2}. Indeed, the prospects of genuine multi-frequency GW observations \cite{sesa16} will greatly increase our chances of testing the existence of such an elusive population of black holes.


\acknowledgements
A.S. acknowledges financial support provided under the European Union’s H2020 ERC Consolidator Grant ``Binary Massive Black Hole Astrophysics'' (B Massive, Grant Agreement no. 818691). E.~B. acknowledges financial support provided under the European Union's H2020 ERC Consolidator Grant ``GRavity from Astrophysical to Microscopic Scales'' grant agreement no. GRAMS-815673. F.H. and M.D. acknowledge funding from
MIUR under the grant PRIN 2017-MB8AEZ. This work was supported by the EU Horizon 2020 Research and Innovation Programme under the Marie Sklodowska-Curie Grant Agreement No. 101007855.

\bibliography{biblio_pbh}

\end{document}